\documentclass[twocolumn]{aastex63}

\newcommand{\HIIs}{H\,{\sc ii}\,}

\newcommand{\NeIIIs}{{[Ne\,{\sc iii}]\,}}

\newcommand{\OIII}{{[O\,{\sc iii}]}}
\newcommand{\OIIIs}{{[O\,{\sc iii}]\,}}

\newcommand{\OIIs}{{[O\,{\sc ii}]\,}}

\newcommand{\FeII}{{[Fe\,{\sc ii}]\,}}

\received{July 16, 2021}
\accepted{July 28, 2021}
\submitjournal{ApJL}

\shorttitle{Mapping metallicity of outflows and inflows in Mrk~1486}
\shortauthors{Cameron et al.}
\graphicspath{{./}{figures/}}

\begin{document}

\title{The DUVET Survey: Direct $T_e$-based metallicity mapping of metal-enriched outflows and metal-poor inflows in Mrk~1486}

\correspondingauthor{Alex J. Cameron}
\email{alex.cameron@physics.ox.ac.uk}

\author[0000-0002-0450-7306]{Alex J. Cameron}
\affil{Department of Physics, University of Oxford, Denys Wilkinson Building, Keble Road, Oxford, OX1 4RH, UK}
\affil{School of Physics, The University of Melbourne, Parkville, VIC 3010, Australia}
\affil{ARC Centre of Excellence for All Sky Astrophysics in 3 Dimensions (ASTRO 3D), Australia}

\author{Deanne B. Fisher}
\affiliation{Centre for Astrophysics and Supercomputing, Swinburne University of Technology, Hawthorn, Victoria 3122, Australia}
\affil{ARC Centre of Excellence for All Sky Astrophysics in 3 Dimensions (ASTRO 3D), Australia}

\author{Daniel McPherson}
\affiliation{Centre for Astrophysics and Supercomputing, Swinburne University of Technology, Hawthorn, Victoria 3122, Australia}
\affil{ARC Centre of Excellence for All Sky Astrophysics in 3 Dimensions (ASTRO 3D), Australia}

\author{Glenn G. Kacprzak}
\affiliation{Centre for Astrophysics and Supercomputing, Swinburne University of Technology, Hawthorn, Victoria 3122, Australia}
\affil{ARC Centre of Excellence for All Sky Astrophysics in 3 Dimensions (ASTRO 3D), Australia}

\author{Danielle A. Berg}
\affil{Department of Astronomy, The University of Texas at Austin, 2515 Speedway, Stop C1400, Austin, TX 78712, USA}

\author{Alberto Bolatto}
\affil{Department of Astronomy, University of Maryland, College Park, MD 20742, USA}

\author[0000-0002-0302-2577]{John Chisholm}
\affil{Department of Astronomy, The University of Texas at Austin, 2515 Speedway, Stop C1400, Austin, TX 78712, USA}

\author{Rodrigo Herrera-Camus}
\affil{Departamento de Astronomía, Universidad de Concepción, Barrio Universitario, Concepción, Chile}

\author[0000-0003-2377-8352]{Nikole M. Nielsen}
\affiliation{Centre for Astrophysics and Supercomputing, Swinburne University of Technology, Hawthorn, Victoria 3122, Australia}
\affil{ARC Centre of Excellence for All Sky Astrophysics in 3 Dimensions (ASTRO 3D), Australia}

\author[0000-0002-7187-8561]{Bronwyn Reichardt Chu}
\affiliation{Centre for Astrophysics and Supercomputing, Swinburne University of Technology, Hawthorn, Victoria 3122, Australia}
\affil{ARC Centre of Excellence for All Sky Astrophysics in 3 Dimensions (ASTRO 3D), Australia}

\author[0000-0001-9719-4080]{Ryan J. Rickards Vaught}
\affil{Center for Astrophysics and Space Sciences, Department of Physics, University of California, San Diego, 9500 Gilman Drive, La Jolla, CA 92093, USA}

\author{Karin Sandstrom}
\affil{Center for Astrophysics and Space Sciences, Department of Physics, University of California, San Diego, 9500 Gilman Drive, La Jolla, CA 92093, USA}

\author{Michele Trenti}
\affil{School of Physics, The University of Melbourne, Parkville, VIC 3010, Australia}
\affil{ARC Centre of Excellence for All Sky Astrophysics in 3 Dimensions (ASTRO 3D), Australia}



\begin{abstract}

We present electron temperature ($T_e$) maps for the edge-on system Mrk~1486, affording ``direct-method'' gas-phase metallicity measurements across $5.\!\!^{\prime\prime}8$ (4.1 kpc) along the minor axis and $9.\!\!^{\prime\prime}9$ (6.9 kpc) along the major axis.
These maps, enabled by strong detections of the \OIIIs$\lambda$4363 auroral emission line across a large spatial extent of Mrk~1486, reveal a clear negative minor axis $T_e$ gradient in which temperature decreases with increasing distance from the disk plane.
We find that the lowest metallicity spaxels lie near the extremes of the major axis, while the highest metallicity spaxels lie at large spatial offsets along the minor axis.
This is consistent with a picture in which low metallicity inflows dilute the metallicity at the edges of the major axis of the disk, while star formation drives metal-enriched outflows along the minor axis.
We find that the outflow metallicity in Mrk~1486 is 0.20 dex (1.6 times) higher than the average ISM metallicity, and more than 0.80 dex (6.3 times) higher than metal-poor inflowing gas, which we observe to be below 5~\% $Z_\odot$.
This is the first example of metallicity measurements made simultaneously for inflowing, outflowing, and inner disk ISM gas using consistent $T_e$--based methodology.
These measurements provide unique insight into how baryon cycle processes contribute to the assembly of a galaxy like Mrk~1486.

\end{abstract}

\keywords{galaxies: abundances --- 
galaxies: evolution --- galaxies: starburst --- ISM: jets and outflows}


\section{Introduction} \label{sec:intro}

The cycle of baryons into galaxies via accretion and back out via outflows is one of the most important regulators of galaxy evolution \citep{SomervilleDave15, Tumlinson17}.
Metal enrichment during this cycle is a critical element of galaxy assembly.
Simulations predict inflows of metal-poor gas along the major axis of the galaxy, while metal-enriched outflows are ejected along the minor axis \citep{Nelson19, Peroux20, Kim20, Mitchell20}.
However, this picture remains poorly constrained by observations.

Gas-phase metallicities in star-forming galaxies are widely derived from emission line ratios (e.g. \citealt{MaiolinoMannucci19, Kewley19_review}).
Although widely used,  strong-line methods to measure metallicity are sensitive to assumptions surrounding the ionization conditions and relative abundance ratios. This approach carries large systematic uncertainties that are difficult to quantify \citep[e.g.][]{KewleyEllison08, Berg11}.
Nevertheless, metallicity measurements have been compiled for large samples of galaxies with resultant scaling relations offering indirect evidence of metal-poor inflows and metal-enriched outflows \citep{Tremonti04, Dalcanton04, Mannucci10, Kacprzak16, Sanders20_MZR}.

Variations in the ionization conditions of halo gas at large separations from the inner disk make it challenging to reliably map metallicity throughout the halo with strong-line methods.
Quasar absorption lines can be used to measure line-of-sight circumgalactic medium (CGM) metallicity, although suitable sightlines are rare and these samples can only be assembled in a statistical way. Currently there is no consensus as to whether the CGM metallicity varies as a function of azimuthal angle from major to minor axis as observational evidence is mixed \citep{Peroux16, Pointon19, Kacprzak19, Lehner20, Lundgren21, Wendt21}.

\citet{Chisholm18} directly measured outflow metallicities from UV absorption lines for a small sample of local galaxies, finding these to be metal-enriched relative to emission line measurements of the interstellar medium (ISM) metallicity.
However, systematic offsets are widely observed between different metallicity measurement techniques \citep[e.g.][]{KewleyEllison08, MaiolinoMannucci19}. Moreover, it is difficult to interpret how the geometry of the gas may affect the metal loading of the outflows. 

The gold standard for emission-line abundance studies is the so-called ``direct method'' in which metallicity is determined using an electron temperature ($T_e$) measurement \citep{PerezMontero17}.
$T_e$ in \HIIs regions reflects the balance of heating from the ionization source and radiative cooling which is highly metallicity dependent. $T_e$ can be measured via the ratio of a so-called ``auroral'' emission line (most commonly \OIIIs$\lambda$4363) to a strong line. Auroral lines are, however, 50-100$\times$ fainter than strong lines.
Accordingly, the direct method is very challenging to apply, even at low redshift, and spatially resolved $T_e$ measurements remain rare \citep{Berg13, Berg20, Li13, Croxall15, Croxall16, Ho19, Leung20, Cameron21}.

In this letter we map direct method abundance variations, using \emph{Keck}/KCWI observations of edge-on starburst galaxy Mrk~1486, along both the major and minor axes, providing unique constraints simultaneously on the metallicity of inflowing and outflowing gas, as well as the average ISM. These abundance measurements, derived from a single self-consistent method, represent a powerful new method for studying metal enrichment throughout the baryon cycle.

\section{Observations and Data Reduction} \label{sec:data}

\begin{figure*}
    \centering
    \includegraphics[width=\textwidth]{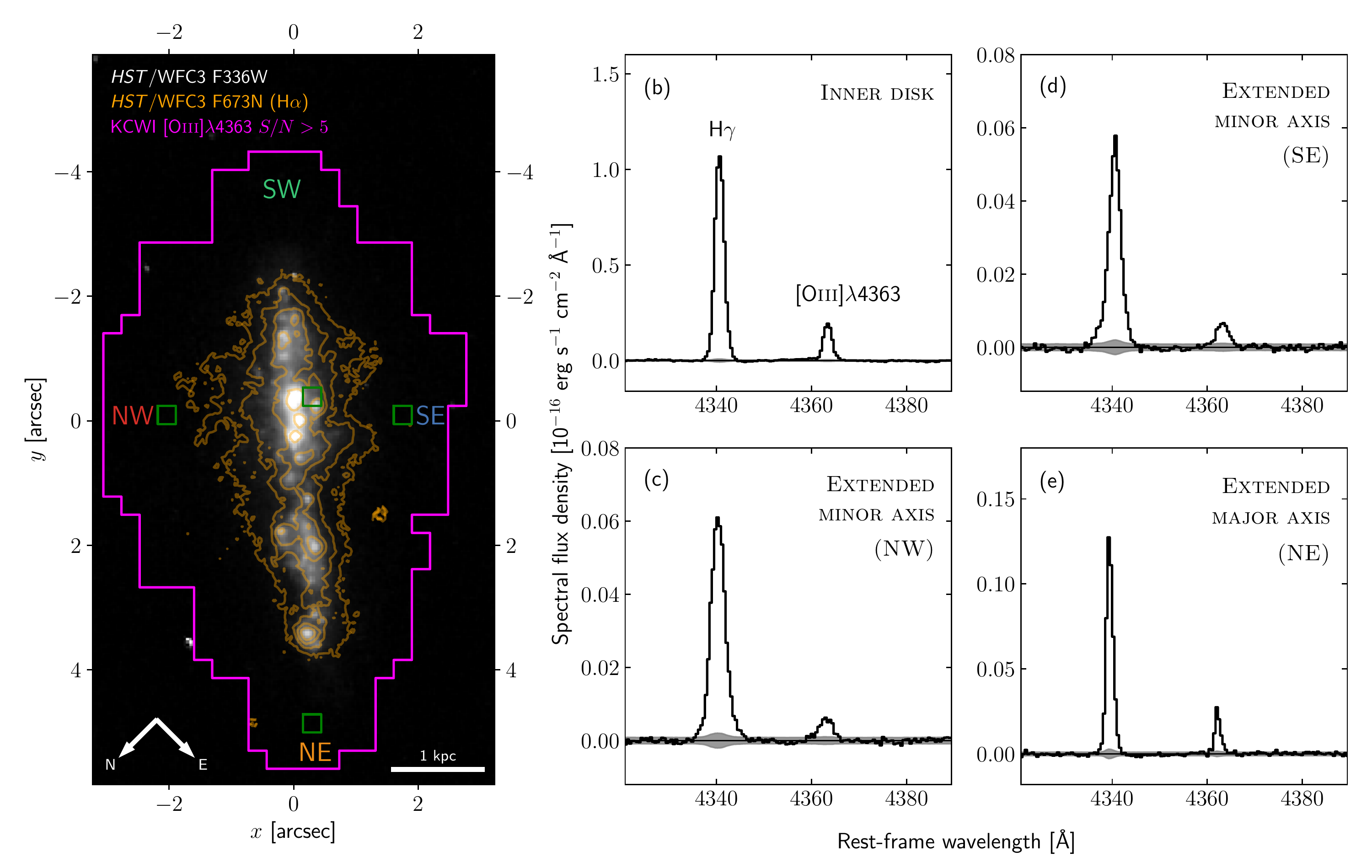}
    \caption{
    Our \OIIIs$\lambda$4363 spatial coverage extends far beyond the starlight observed in Mrk~1486.
    \textit{Left panel:} Magenta footprint shows the contiguous region in which we observe $S/N_{\lambda 4363}>5$ per spaxel for the auroral line in our KCWI observations, compared to $U$-band flux from \emph{HST}/WFC3 imaging (grey scale image).
    Orange contours show extended H$\alpha$ emission from \emph{HST}/WFC3 F673N narrow-band imaging with levels of [0.69, 1.7, 4.3, 11] $\times$10$^{-19}$ erg s$^{-1}$ cm$^{-2}$.
    \textit{Panels (b-e):} Zoomed in view of \OIIIs$\lambda$4363 in continuum-subtracted spectra at various spatial locations across the KCWI data cube (corresponding to green squares in the left panel), demonstrating our unambiguous detection of the auroral line, even at large separations. Grey shaded interval indicates the 1-$\sigma$ error spectrum.
    }
    \label{fig:1}
\end{figure*}

Mrk~1486 is a star-forming galaxy at $z=0.033841$ ($r$-band magnitude $m_r = 16.81$) with stellar mass log($M_*/M_\odot) = 9.3\pm0.2$ and star-formation rate SFR $= 3.6\pm0.7~ M_\odot$ yr$^{-1}$ \citep{Chisholm18}.
In \emph{HST}/WFC3 imaging, Mrk~1486 has morphology consistent with a disk-like galaxy that is oriented almost directly edge-on (Figure~\ref{fig:1}).
Narrow-band imaging shows extended H$\alpha$ emission along the minor axis and previous studies have shown that this is consistent with bipolar outflows \citep{Duval16}.
The metallicity reported for Mrk~1486 ($12+\text{log}(O/H)=7.80$; \citealt{Ostlin14}) is very low for its stellar mass, consistent with recent accretion of metal-poor gas. 
Together, these suggest Mrk~1486 is a unique target for studying baryon cycle gas flows.

Integral field unit (IFU) observations of Mrk~1486 were carried out on 2020 March 22nd with \emph{Keck}/KCWI as part of the \underline{D}eep near-\underline{UV} observations of \underline{E}ntrained gas in \underline{T}urbulent galaxies (DUVET) survey (Fisher et al., \textit{in prep.}) in good seeing conditions ($\sim0.\!\!^{\prime\prime}7$ at 5000 \AA{}).
The large IFU slicer setting was used giving a spatial sampling of $0.\!\!^{\prime\prime}29 \times 1.\!\!^{\prime\prime}35$ over a $20^{\prime\prime} \times 33^{\prime\prime}$ field-of-view.
Two configurations of the blue medium-dispersion grating (BM) were used
(``blue'' central wavelength: 4180 \AA{}; ``red'' central wavelength: 4850 \AA{})
providing continuous spectral coverage from 3731 \AA{} to 5284 \AA{}\footnote{The two settings overlap between 4408 \AA{} and 4627 \AA{}} with spectral resolution $R\sim2000$, affording coverage of emission lines from \OIIs$\lambda\lambda$3726, 3729 to \OIIIs$\lambda$5007.
Total integration for the red grating was 2260 s across seven long exposures ($6\times300$ s and $1\times400$ s) and two short exposures ($2\times30$ s). The blue grating setting received $7\times300$ s exposures for 2100 s of total integration.

The data were reduced with the KCWI Data Extraction and Reduction Pipeline v1.1.0.
Long exposures in the red grating setting saturated \OIIIs$\lambda$5007 in some spaxels. Saturated spaxels were identified based on the \OIIIs$\lambda$5007/$\lambda$4959 ratio (which should have a constant value of $R_{\lambda5007/\lambda4959}=3$) and spectral pixels proximal to \OIIIs$\lambda$5007 were replaced by measurements from the shorter exposures (McPherson et al. \textit{in prep.}).

Exposures were spatially aligned using a linear interpolation of the H$\gamma$ flux which was detected and unsaturated in all frames (short and long, red and blue). Sky subtraction was performed within the pipeline using an off-object sky frame taken immediately before or after a set of exposures.

\section{Spectral fitting} \label{sec:methods}

\subsection{Continuum subtraction}

We perform four-moment fits to the stellar continuum
with model spectra from the Binary Population and Spectral Synthesis code (BPASS, v2.2.1; \citealt{Eldridge17, StanwayEldridge18}\footnote{BPASS model spectra can be easily handled with the purpose built python package \texttt{hoki} (https://github.com/HeloiseS/hoki; \citealt{HOKI20})})
for both grating settings of each spaxel with \texttt{pPXF} \citep{pPXF17}, adopting the `135\_300' IMF (refer to Table 1 in \citealt{StanwayEldridge18}).
Since we detect many faint emission lines in our KCWI data, we first run a continuum fit on the summed global spectrum of Mrk~1486 in which the iterative sigma-clipping approach of \citet{Cappellari02}\footnote{This is contained within the \texttt{pPXF} package and can be used by setting the \texttt{clean=True} keyword in the main \texttt{pPXF} routine.} is employed to mask out these faint emission features.
We apply these masks to individual spaxels during subsequent fitting to minimise the impact of faint emission features on the template fitting.

\subsection{Emission line fitting}

Emission line fits are performed on the continuum-subtracted data for each spaxel according to the following procedure:
velocity (or redshift, $z$) and velocity dispersion ($\sigma$) is obtained by simultaneously fitting single-component Gaussian profiles to the H$\beta$, H$\gamma$, and \OIIIs$\lambda\lambda$4959, 5007, emission features in the red setting using a $\chi^2$-minimisation procedure.
Treating these values as fixed for that spaxel, fluxes are then fit individually for each emission line, using a 30 \AA{} sub-interval of the complete spectrum centred on the expected centroid of the line at the best-fit redshift.
We fit the \OIIs$\lambda\lambda$3726, 3729, \NeIIIs$\lambda$3869, H$\delta$, H$\gamma$, and \OIIIs$\lambda$4363 lines in the blue grating setting, and the H$\gamma$, \OIIIs$\lambda$4363, H$\beta$, \OIIIs$\lambda$4959, \OIIIs$\lambda$5007 lines in the red grating setting.\footnote{At metallicities above $12+\text{log}(O/H)\gtrsim8.4$, \OIIIs$\lambda$4363 can be contaminated by emission from \FeII$\lambda$4360. Mrk~1486 is, however, much lower metallicity and shows no evidence of such contamination.}
\OIIs$\lambda\lambda$3726, 3729 doublet lines are fit simultaneously, since they are partially blended at this resolution; no restrictions are imposed on the flux ratio.
Additionally, we fit the \OIIIs$\lambda$5007 and \OIIIs$\lambda$4959 fluxes independently to verify that the saturation of \OIIIs$\lambda$5007 has been appropriately corrected for in the data reduction (see Section~\ref{sec:data}). We corrected for dust extinction based on the H$\gamma$/H$\beta$ flux ratio observed in the red grating setting, assuming an intrinsic Balmer decrement of $f_{H\gamma}/f_{H\beta}=0.468$ \citep{DopitaSutherland03, Groves12} and a \citet{Cardelli89} extinction law with $R_V=3.1$. We found a peak extinction of $A_V = 0.96\pm0.12$ ($E(B-V)=0.31\pm0.04$) on the disk. At large separations, $A_V\approx0$ was commonly observed, suggesting little extincting dust is present in the outflows and inflows, perhaps due to the low metallicity of the system.

Line flux uncertainties are estimated using a bootstrapping method in which the observed spectrum is perturbed at each spectral pixel by a normal distribution with standard deviation derived from the KCWI variance cube. We measure fluxes for 100 such synthetic spectra and the standard deviation of the resulting flux distribution is adopted as the 1-$\sigma$ flux uncertainty.
Additionally, uncertainties from the reddening correction were propagated through to the line ratios and derived properties presented throughout the remainder of the paper.


\section{Results} \label{sec:results}

\begin{figure*}
    \centering
    \includegraphics[width=0.9\textwidth]{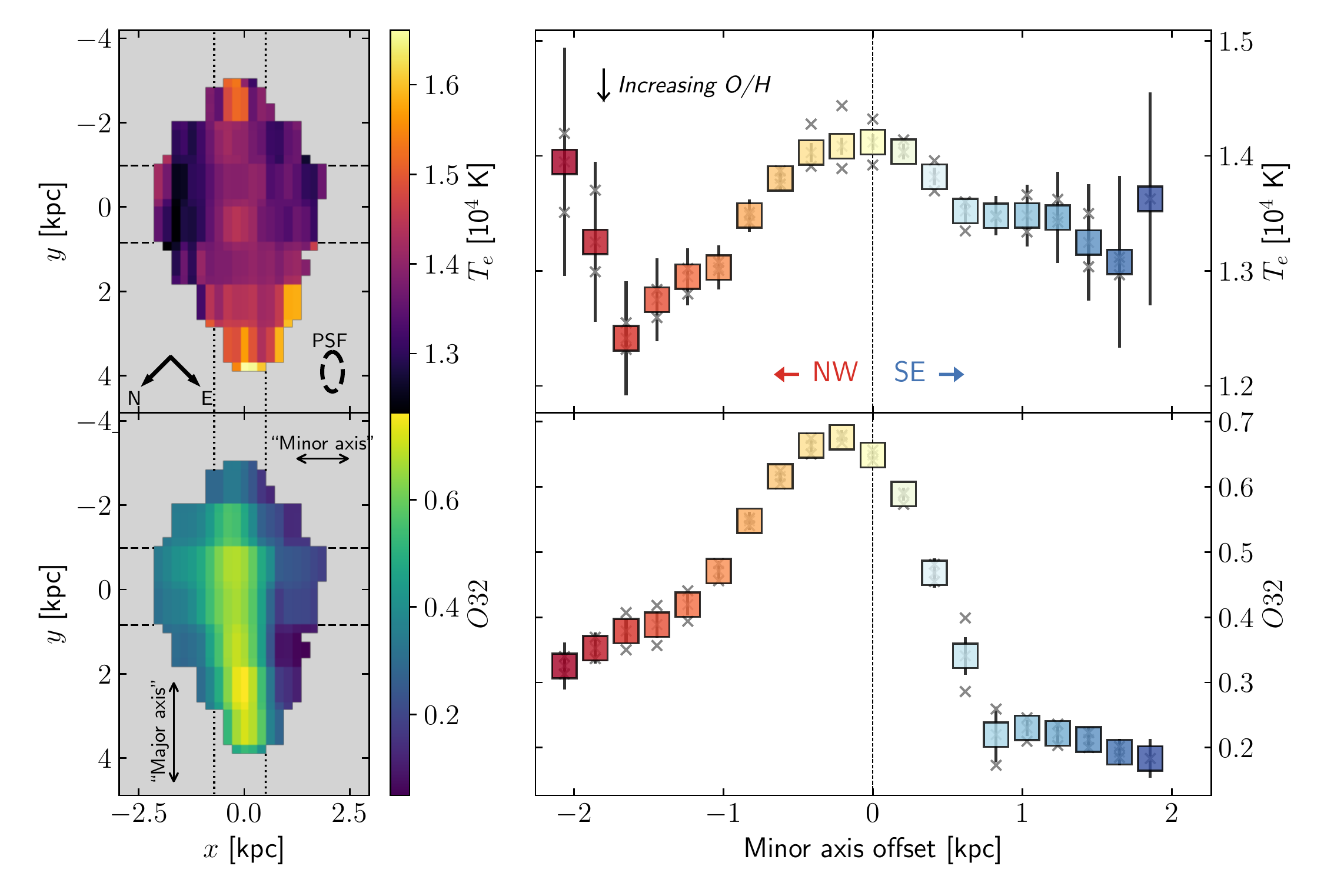}
    \includegraphics[width=0.9\textwidth]{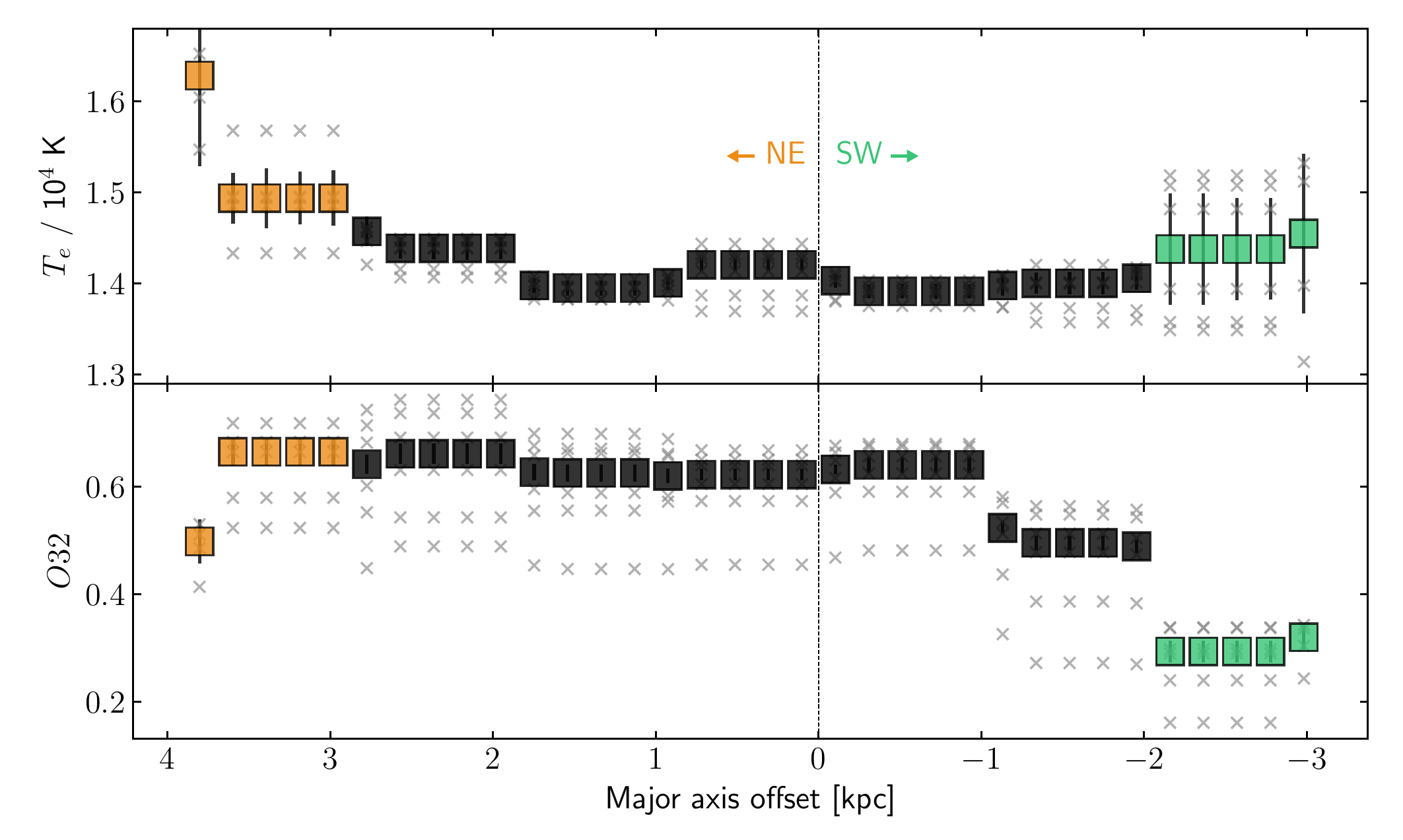}
    \caption{
    \textit{Top left:} Maps of $T_e$ (upper) and $O32$ (lower) for Mrk~1486. The orientation is the same as for Figure~\ref{fig:1} and the spatial coverage matches the magenta footprint in that Figure.
    \textit{Top right:} $T_e$ and $O32$ profile against minor axis offset for the region bounded by the dashed lines in the top left panel. Grey crosses indicate individual values, while large squares are median values with error bars depicting the median uncertainty of values in that bin.
    Negative minor axis gradients are observed for both $T_e$ and $O32$.
    \textit{Bottom:} As for top right, except showing the major axis profile (region bounded by the dotted lines in the top left panel). Note that the major axis sampling is coarser due to the shape of the KCWI spaxels. Grey crosses again correspond to individual points in the final resampled $0.\!\!^{\prime\prime}29 \times 0.\!\!^{\prime\prime}29$ science cube. These points have a large covariance with adjacent points along this major axis direction.
    The color coding of minor and major axis profiles corresponds to that used in Figure~\ref{fig:model_grids}.
    }
    \label{fig:minor_gradient}
\end{figure*}

\subsection{Electron temperature}
\label{sub:te_map}

Our \emph{Keck}/KCWI observations afford detections of the critical \OIIIs$\lambda4363$ auroral line at $S/N>5$ per spaxel over a spatial extent of $9.\!\!^{\prime\prime}9$ (6.9 kpc) along the major-axis and $5.\!\!^{\prime\prime}8$ (4.1 kpc) along the minor-axis, well beyond $R_{90}$ (Figure~\ref{fig:1}).
We derive an electron temperature ($T_e$) map for Mrk~1486 from the \OIIIs$\lambda$4363 / $\lambda$5007 emission line ratio according to the relation set out by \citet{Nicholls20}, with resultant temperatures in the range $1.21\pm0.09 \leq T_e / 10^4~\text{K} \leq 1.81\pm0.08$ (Figure~\ref{fig:minor_gradient} top left panel).
The edge-on orientation of Mrk~1486 enables us to derive $T_e$ profiles along and above the disk plane with consistent methodology, providing unique direct-method metallicity measurements of inflowing and outflowing gas.

The upper right panel of Figure~\ref{fig:minor_gradient} shows our derived minor axis $T_e$ profile for a $2.\!\!^{\prime\prime}61$ wide strip centred on the peak KCWI white light flux.
$T_e$ is seen to peak on the plane of the disk at  $T_e / 10^4\text{ K} = 1.41 \pm 0.01$ and decreases in both directions.
This decrease is most significant in the Northwest (NW; red) direction, dropping by $(0.17\pm0.05)\times10^4$ K to $T_e / 10^4\text{ K} = 1.24 \pm 0.05$ at $R_{\text{minor}} = -1.6$ kpc.
In the Southeast (SE; blue) direction we instead measure a decrease of $(0.10\pm0.07)\times10^4$ K to $T_e / 10^4\text{ K} = 1.31 \pm 0.07$ at $R_{\text{minor}} = 1.6$ kpc.
In both directions, the gradient in $T_e$ reverses in the final resolution elements. 

Our derived major axis $T_e$ profile is shown in the lower panels of Figure~\ref{fig:minor_gradient}. The highest temperature values preferentially reside at larger radius, consistent with a negative radial metallicity gradient.

Recent studies have shown that metallicity measurements derived from only a single temperature measurement can result in misleading metallicity trends if the ionization conditions of the emitting \HIIs regions are not constant \citep{Yates20, Cameron21}. Figure~\ref{fig:minor_gradient}, therefore, shows the $O32$\footnote{$O32$ = log (\OIIIs$\lambda$5007 / \OIIs$\lambda\lambda$3726, 3729)} profile along the minor and major axes. This ratio is a commonly used ionization parameter (log $U$) diagnostic.
There is a negative $O32$ minor axis gradient, consistent with the fiducial assumption that ionization parameter will be lower further  from the disk.
Since our DUVET observations of Mrk~1486 yield an auroral line ratio only for \OIII, in the next sections we use predictions from photoionization modelling to explore the contribution of ionization parameter to the observed temperature gradient, and ultimately infer metallicity variations.

\subsection{{\sc Mappings} model grids}
\label{sub:model_grids}

We compare observed $R_{O3} =$ \OIIIs$\lambda$4363 / $\lambda$5007 and $O32$ line ratios from Mrk~1486 to predictions from {\sc Mappings V} model grids (A. D. Thomas; private communication).
These models were computed with {\sc Mappings V} \citep{Sutherland17} for a STARBURST99 continuous star-formation model, ensuring a fully populated IMF, \citep{Leitherer14} over a range of nebular metallicity, ionization parameter, and ISM pressure values.

We find that while the $R_{O3}$ ratio is dependent on pressure in high metallicity systems, the effect is minimal below $12+\text{log}(O/H)\lesssim8.3$, which our $R_{O3}$ measurements strongly favour for Mrk~1486. What little pressure dependence  $R_{O3}$ exhibits at low metallicity is strongest at high pressure.
However, our measurements of the \OIIs$\lambda$3729/$\lambda$3726 doublet ratio, which can be used as an effective pressure diagnostic \citep{Kewley19_pressure}, strongly favour low pressure (log$(P/k)\lesssim6.6$) throughout Mrk~1486. Thus, we conclude that our observed temperature variations are not caused by pressure variations. We measure a median density of $n_e{\text{(O{\sc ii})}} = 100$ cm$^{-3}$ and infer a median pressure of log$(P/k)=6.0$.
For the remainder of the paper, we fix the pressure to log$(P/k) = 5.8$ in our model grids.\footnote{The model grids were computed for 12 pressure values, of which log$(P/k) = 5.8$, 6.2 are nearest our observed pressure. Since the effect of pressure on the $R_{O3}$ ratio is so small at low metallicity, adopting either of these two values ultimately makes no difference to our results.}

\begin{figure*}
    \centering
    \includegraphics[width=0.9\textwidth]{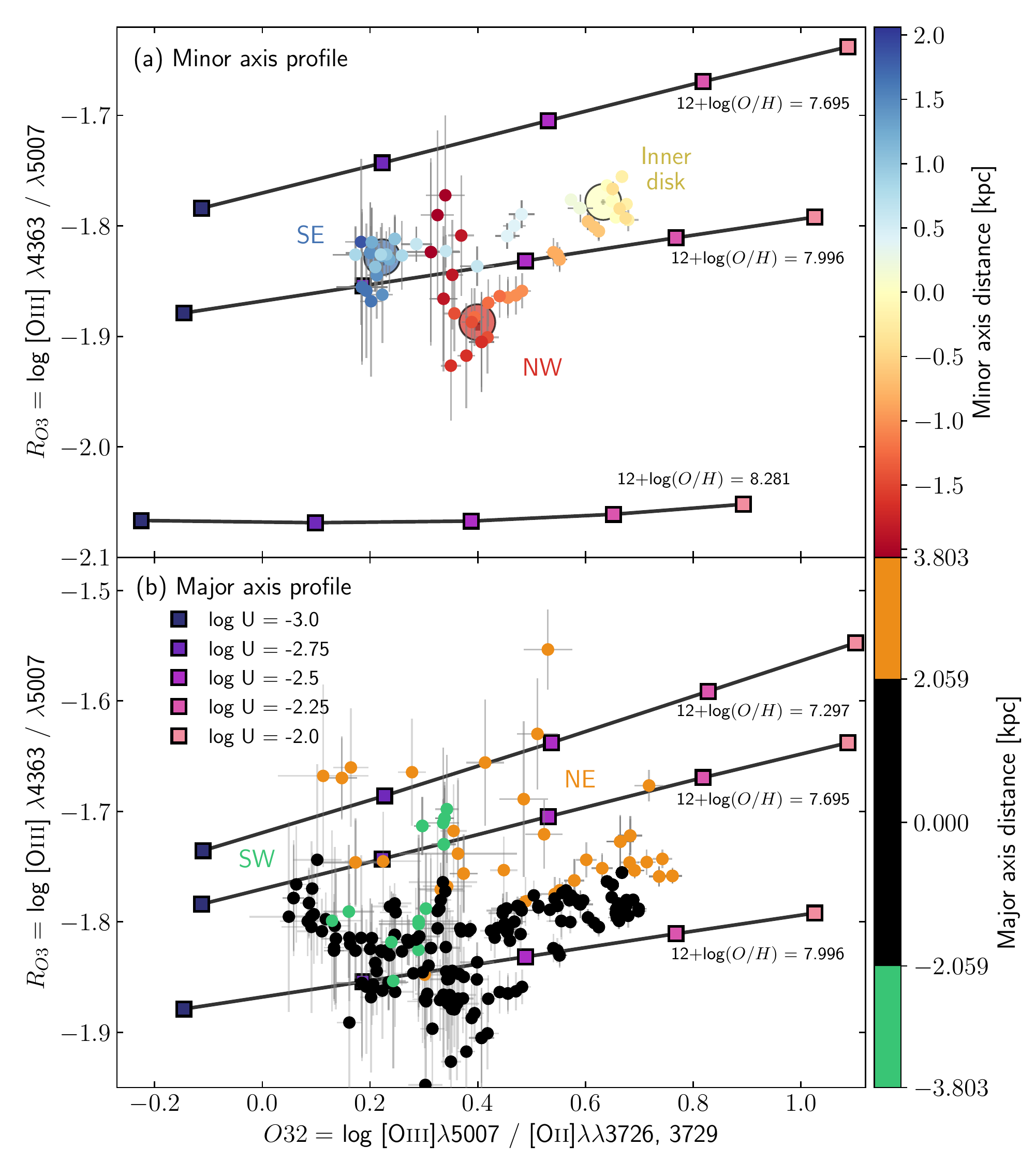}
    \caption{
    Observed $R_{O3}$ and $O32$ line ratios for individual spaxels in Mrk~1486 compared to predictions from {\sc Mappings} grids.
    Square markers show model predictions color coded by model log $U$. Solid lines connect grid points of the same metallicity.
    Observed points (circles) are from individual spaxels with the color bar showing the projected offset from the core of Mrk~1486.
    The colored compass labels relate the points back to Figure~\ref{fig:1}.
    \textit{Panel (a):} Spaxels shown lie in a $2.\!\!^{\prime\prime}61$ wide aperture along the minor-axis (same minor axis selection as Figure~\ref{fig:minor_gradient}).
    We observe a temperature decrease in both directions. In the NW direction (red) in particular, this appears to be consistent with increasing metallicity from the core to $R_{\text{minor}}=-1.6$ kpc.
    Large circles depict values derived for three $0.\!\!^{\prime\prime}87 \times 1.\!\!^{\prime\prime}45$ apertures along the minor axis. One in the core (yellow), and two offset along the minor axis (centred on highlighted spaxels from Figure~\ref{fig:1}).
    \textit{Panel (b):} 
    All observed points with $S/N_{\lambda4363}>5$ with a discrete color bar indicating major axis offset. Black points are within the 90\% $i$-band flux radius ($R_{90}$).
    Green and orange points lie beyond $R_{90}$ in the SW and NE directions respectively.
    Of the lowest metallicity ($12+\text{log}(O/H)<7.695$) points, 94~\% (17/18)  lie beyond $R_{90}$, and the most distant major axis points all fall into this regime.
    }
    \label{fig:model_grids}
\end{figure*}

Figure~\ref{fig:model_grids} shows how predictions for the $R_{O3}$ auroral line ratio (which traces $T_e$) and the $O32$ ratio (which traces log $U$) compare to the observed values along the minor and major axes of Mrk~1486. 
For visual clarity, we plot the models only for four values of metallicity,
each at five different values of ionization parameter.
The left panel of Figure~\ref{fig:schematic} then shows a metallicity map for Mrk~1486 derived from interpolation of these grids, performed with a $\chi^2$-minimisation procedure based on the $R_{O3}$ and $O32$ line ratios.
The lowest metallicities we infer have $12+\text{log}(O/H)\lesssim7.297$, lying beyond the low metallicity extreme of our model grids. Inferring oxygen abundances in this extremely low metallicity regime is very susceptible to modelling uncertainties and has extremely few empirical constraints from observations \citep[e.g.][]{Senchyna19, StanwayEldridge19}. 
Direct $T_e$ measurements from a lower ionization zone (e.g. $T_e$(O{\sc ii}))
are required to robustly determine the oxygen abundance of most metal-poor spaxels in Mrk~1486.
We instead adopt $12+\text{log}(O/H)\lesssim7.297$ as an upper limit in these cases for remainder of this analysis.

\subsection{Minor axis metallicity variation: metal-enriched outflows}

In panel (a) of Figure~\ref{fig:model_grids} we show observed minor axis points from a $2.\!\!^{\prime\prime}61$ wide region centred on the  peak of the white light flux, with the color bar indicating the projected offset.
Particularly notable is the NW (red) measurements decreasing steeply toward lower $R_{O3}$. This suggests that the metallicity increases out to a minor axis offset of  $R_{\text{minor}} = -1.6$ kpc in this direction.

\begin{figure*}
    \centering
    \includegraphics[width=0.9\textwidth]{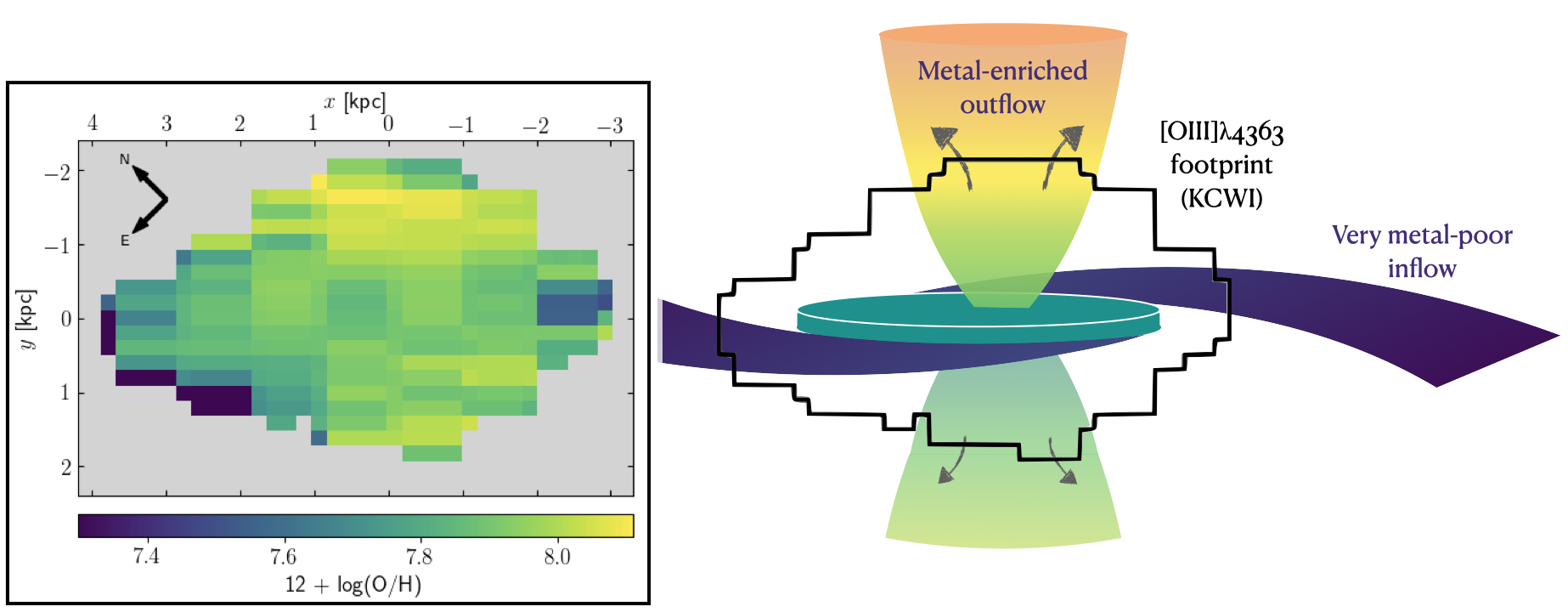}
    \caption{
    \textit{Left panel:} map of inferred metallicities (see \S~\ref{sub:model_grids}).
    The highest metallicity regions are offset from the disk along the minor axis. The lowest metallicity points are generally at large separations along the major axis.
    \textit{Right panel:} Schematic diagram depicting qualitative metallicity differences from different gas flow components. Black footprint shows the extent of the metallicity map from KCWI observations.
    Our high metallicity points along the minor axis can be explained by the presence of outflow cones driving out metal-enriched gas.
    The asymmetry observed along the minor axis may indicate that these outflows are inhomogeneous.
    Our lowest metallicity points at the extremes of the major axis extend beyond the 90~\% $i$-band flux radius of Mrk~1486 and consequently may see a much larger dilution effect from the inflow of metal-poor gas.
    Given we measure metallicity as low as $12+{\text{log}}(O/H)\approx7.3$ (upper limit; refer to \S~\ref{sub:model_grids}) at these extremes, we infer that the source of the inflowing gas may be extremely metal poor.
    }
    \label{fig:schematic}
\end{figure*}

The increase in minor-axis metallicity is clearly seen in the metallicity map (Figure~\ref{fig:schematic}). The highest metallicities we observe in Mrk~1486 are offset from the disk plane along the minor axis, particularly in this NW direction. This is consistent with a scenario in which star formation activity drives metal-enriched outflows (see Figure~\ref{fig:schematic} right panel), which has been widely invoked to explain metallicity scaling relations \citep[e.g.][]{Tremonti04, Mannucci10, Sanders20_MZR}. The maximum metallicity measured along the minor axis ($12+\text{log}(O/H)=8.10^{+0.07}_{-0.06}$ at $R_{\text{minor}} = -1.6$ kpc) is 1.6 times (0.20 dex) higher than the luminosity-weighted average metallicity of the inner disk.

\citet{Chisholm18} used UV absorption line measurements from \emph{HST}/COS to measure outflow metallicities for a sample of low-redshift galaxies including Mrk~1486. Their data yield a much higher outflow metallicity for Mrk~1486 of $Z_{\text{outflow}}/Z_\odot=0.56\pm0.03$ ($12+\text{log}(O/H)=8.44$ assuming a solar value of $12+\text{log}(O/H)_\odot=8.69$).

Using our measurement of the ISM metallicity, this yields an outflow enrichment factor of $Z_{\text{outflow}}/Z_{\text{ISM}}=3.5$ for Mrk~1486.\footnote{\citet{Chisholm18} use the ISM metallicity determination from \citet{Ostlin14} which results in a larger quoted value of $Z_{\text{outflow}}/Z_{\text{ISM}}=4.3$ for Mrk~1486}
While this value is substantially higher than ours, we note that different metallicity measurement techniques are prone to systematic offsets and values obtained via different methods are difficult to compare \citep[e.g.][]{KewleyEllison08, MaiolinoMannucci19}. Furthermore, the outflow metallicity measured by \citet{Chisholm18} is derived from kinematically offset absorption lines observed for the $2.\!\!^{\prime\prime}5$ aperture of COS, centred on the plane of the disk. This measurement probes gas that is flowing along the line-of-sight towards us. Our DUVET observations instead probe outflowing gas along a projected spatial offset. It is not simple to understand how these differences in geometry would affect the metallicity.

The metallicity increase is milder in the SE direction, only reaching a maximum value of $12+\text{log}(O/H)=8.02^{+0.08}_{-0.14}$ at $R_{\text{minor}} = 1.7$ kpc. This suggests that the bipolar outflows in Mrk~1486 may be asymmetric, and that the NW outflow is perhaps more enriched at this point in the evolution of Mrk~1486.
In both minor axis directions, we observe that the metallicity decreases for the few spaxels at the largest offsets (see also Figure~\ref{fig:minor_gradient} upper right panel).
This could be an indication of inhomogeneity within an outflow lobe, suggesting a clumpy structure.
However, we strongly caution against over interpreting these few data points. These are our faintest spaxels and the \OIII$\lambda$4363 peak line flux is only $\sim$10$^{-19}$ erg s$^{-1}$ cm$^{-2}$ \AA{}$^{-1}$. At these low fluxes, systematic errors from sky subtraction may be important, which are not currently included in our error bars. Hence we stop short of drawing strong conclusions from these points without deeper data.


\subsection{Major axis metallicity variation: metal-poor inflows}

In panel (b) of Figure~\ref{fig:model_grids}, all observed spaxels are plotted with a discrete colorbar distinguishing points within the major-axis 90\% $i$-band flux radius (black) from those beyond it (orange and green).
We find that spaxels nearer to the core are generally observed to have higher metallicity.
Almost 95~\% (17/18) of points consistent with very low metallicity gas ($12+\text{log}(O/H)<7.695$) are situated outside $R_{90}$.
For an edge-on galaxy, a line-of-sight measurement at a given projected radius along the major-axis will sample emission from a range of physical radii from the galaxy center. At low projected radius, observed line-of-sight emission will be dominated by gas from the inner disk, which is likely to have been enriched by recent star formation activity. If a disk is experiencing accretion, at the largest projected radii the emission may arise from a mixture of gas in the outer stellar disk, and dilution from metal-poor inflows. These measured metallicities are thus likely to be over-estimates of the metallicity of accreting gas. 

Near the major axis NE extreme we observe a number of points with $12+\text{log}(O/H)\lesssim7.297$. This upper limit is at least 0.6 dex lower than the ISM metallicity ($Z_{\text{inflow}}/Z_{\text{ISM}}\lesssim0.25$).
Our observed major axis metallicity trend does not stabilise before the coverage cuts off, suggesting that even lower metallicities might be observed at larger radii.
This indicates that the sources of the gas inflow into Mrk~1486 are likely below $\sim$5\%~$Z_{\odot}$.
From our measurements we report that outflowing gas from Mrk~1486 is at least 6.3 times (0.8 dex) more metal rich than inflowing gas, noting that this value represents a lower limit. Direct $T_e$ measurements from a lower ionisation zone (e.g. $T_e$(O{\sc ii})) and improved spatial coverage from deeper data will be important in better constraining this value.


\subsection{Implications for galaxy evolution}

Metal-enriched outflows and metal-poor inflows are frequently invoked as an important mechanism for explaining a number of galaxy observables including scaling relations \citep[e.g.][]{Tremonti04, Dalcanton04, Mannucci10, Sanders20_MZR} and metallicity gradients \citep{Sharda21_primary, Sharda21_MZGR}. The difference between our $Z_{\text{outflow}}/Z_{\text{ISM}}$ and those observed in \cite{Chisholm18} are roughly a factor of 2--3. This is significant compared to differences in analytic predictions of what values of $Z_{\text{outflow}}/Z_{\text{ISM}}$ are required to match observed Mass-Metallicity Relations (MZR), particularly in the mass range of $10^9-10^{10}$~M$_{\odot}$. We note that the difference between these two observations, as well as the asymmetry in $Z_{\text{outflow}}/Z_{\text{ISM}}$ we observe in Mrk~1486, are both reasons motivating the dire need for observations like those reported here to understand basic observables like the MZR.

Metal loss from metal-enriched outflows is predicted to be a critical parameter in determining metallicity gradients.
The model outlined by \citet{Sharda21_primary} quantifies this by the so-called ``yield reduction factor'', $\phi_y$. Larger values of $\phi_y$ imply that metals are mixed efficiently with the ISM prior to ejection, and hence metal retention is higher. This results in steeper metallicity gradients and a larger possible range of gradients.
Adopting mass loading factors $\mu\approx1-4$ \citep{Heckman15,Chisholm18}, our measurement of $Z_{\text{outflow}}/Z_{\text{ISM}}$ from Mrk~1486 yields an estimated value in the range $\phi_y\approx0.8-0.95$ (refer to Appendix A in \citealt{Sharda21_MZGR}).
This predicts galaxies with masses $\sim$10$^{8.5}$ -- 10$^{9.5}$ $M_\odot$ to have steeper gradients than those observed in large IFU surveys \citep[Figure~2 in][]{Sharda21_MZGR}.
On the other hand, models with $\phi_y$ in our measured range reproduce \textit{global} metallicity scaling relations more readily. 
We note, however, that Mrk~1486 is a multiple sigma outlier below the global MZR, and it remains unclear how appropriate these generalized models are for starbursting galaxies.
Finally, we note that our observations probe only the warm ($\sim$10$^4$ K) ionized outflowing phase. Thus, our estimates of the outflow metallicity do not include metals in hotter phases, which may be significant \citep[e.g.][]{Lopez20}.
Any metal enrichment in hot phase outflows would decrease $\phi_y$ for the same mass loading factor and, in the absence of hot phase outflow measurements, the quoted value of $\phi_y$ should be considered as an upper limit.

\section{Discussion}

A basic expectation of the baryon cycle is that gas expelled as outflows will be chemically enriched relative to the inflowing gas that fuels star formation \citep[][and references therein]{Tumlinson17}.
However, practical challenges associated with measuring chemical abundances mean there is an absence of observations directly comparing the abundances in these regimes.

Mrk~1486 presents a unique opportunity to study chemical processing throughout the baryon cycle.
Previous studies have shown that the extended minor-axis emission of Mrk~1486 is consistent with bipolar outflowing nebulae \citep{Duval16}. Similarly, kinematic line-of-sight observations are consistent with outflows \citep{Chisholm15, Chisholm18}.
Mrk~1486 is offset below the local mass-metallicity relation by multiple sigma \citep{Ostlin14, Curti20_MZR}. Considering also the high SFR of Mrk~1486, this offset can be explained by the recent accretion of large amounts of metal-poor gas \citep[e.g.][]{Mannucci10}.
Given the thin disk observed in starlight (Figure~\ref{fig:1}) and well-ordered, disk-like rotation field, it is unlikely Mrk~1486 is undergoing a major merger. We, thus, conclude it is likely experiencing a period of gas accretion that is fuelling a starburst, which is driving bipolar outflows.

Due to the edge-on orientation of Mrk~1486, our DUVET observations map direct method metallicity at all azimuthal angles out to $\sim$2 kpc away from the inner disk, well beyond the $i$-band 90\% flux radius (Figure~\ref{fig:schematic}).
These measurements fit remarkably well with the basic baryon cycle scenario: (1) our lowest metallicity points lie at the major axis extremes (presumably dominated by inflows); (2) our highest metallicity points lie at large offsets along the minor axis nebulae (likely outflows).
Moreover, in mapping abundances with a consistent direct method technique, we are uniquely placed to provide constraints on the degree of chemical enrichment arising from gas processing from inflow to outflow in Mrk~1486.
We measure a factor of 6.3 metallicity increase (0.80 dex) from the major axis extreme to the minor axis extreme.

\citet{Peroux20} found a similar azimuthal metallicity dependence in  TNG50 and EAGLE simulated galaxies at high impact parameters ($R\gtrsim100$ kpc). However, they found this dependence decreases at lower separations and disappears almost entirely within $R\lesssim25$ kpc, in conflict with our measurements of Mrk~1486. There is either a very significant azimuthal mixing process between a few-to-25~kpc, or alternatively the wind material does not extend to the distances probed in the simulations. We note that there is marginal evidence of a decrease in the metallicity at the largest distance along the minor-axis, but this hinges on interpretation of extremely faint emission lines.
An alternate possibility is that the simulations could be not appropriate to describe galaxies like Mrk~1486, which are in a starburst phase.

Previous observational studies on the azimuthal dependence of metallicity in galaxy haloes from quasar absorption line measurements have not reached a clear consensus \citep[e.g.][]{Pointon19, Wendt21}.
However, the requirement for a serendipitous quasar along the line-of-sight means that these measurements must be compiled in a statistical way, averaged across a sample of galaxies.
Furthermore, absorption line measurements tend to be made at larger separations (ranging from $R\sim20-200$ kpc), and low metallicity regions can be masked by high metallicity systems along the line-of-sight where metal absorption is more prominent.
Based on our measurements of Mrk~1486, there is a clear azimuthal metallicity variation within the inner few kpc from the major to the minor axis for this single low metallicity galaxy. More observations like ours are direly needed to determine the extent to which this is representative of the broader galaxy population.

At the extremes of the major axis, we observe a number of points with metallicities below $\sim$5--10\% solar (Figures~\ref{fig:model_grids}~\&~\ref{fig:schematic}). Since \OIIIs$\lambda$4363 emission is biased toward low metallicity gas, if the minor axis nebulae were a mixture of high and low metallicity gas, we would expect to observe low metallicity points, similar to that seen along the major axis.
The absence of these in our observations provide evidence for the lack of minor-axis accretion. Instead, our data are consistent with a co-planar accretion model.
Moreover, we note that inflowing gas along the major axis is likely mixed with disk gas in our observations meaning our measurement is likely an upper limit on the metallicity of inflowing gas.
This suggests the degree of baryon cycle enrichment could be higher than the 0.80 dex quoted above.
Furthermore, it may imply that extremely low metallicity sources of gas can persist even at $z\sim0$.

We conclude by reiterating that Mrk~1486 is in a starburst phase and hence, by definition, not a typical $z\sim0$ galaxy. We therefore advise caution in extending conclusions presented here to the broader galaxy population.
Nevertheless, while observational studies and simulations often focus on ``average'' galaxies, we have highlighted here the power of studying a low metallicity, high SFR galaxy such as Mrk~1486 where the key \OIIIs$\lambda$4363 emission line can be mapped out in detail. Galaxy simulations providing testable predictions for starburst galaxies like Mrk~1486 would provide a useful comparison for single object studies such as this.
Moreover, broader observational studies are required to understand whether findings presented here for Mrk~1486 are common amongst other starbursting galaxies, and to what extent they are representative of local galaxies with lower SFR.



\acknowledgments

We are grateful for conversations with M.~Krumholz and P.~Sharda in preparation of this manuscript. 

The data presented herein were obtained at the W. M. Keck Observatory, which is operated as a scientific partnership among the California Institute of Technology, the University of California and the National Aeronautics and Space Administration. The Observatory was made possible by the generous financial support of the W. M. Keck Foundation.
The authors wish to recognize and acknowledge the very significant cultural role and reverence that the summit of Maunakea has always had within the indigenous Hawaiian community.  We are most fortunate to have the opportunity to conduct observations from this mountain.

This research was supported by the Australian Research Council Centre of Excellence for All Sky Astrophysics in 3 Dimensions (ASTRO 3D), through project number CE170100013. AJC acknowledges support from an Australian Government Research Training Program (RTP) Scholarship and from an Albert Shimmins Postgraduate Writing Up Award through the University of Melbourne. AJC has received funding from the European Research Council (ERC) under the European Union’s Horizon 2020 Advanced Grant 789056 “First Galaxies”. DBF acknowledges support from Australian Research Council (ARC) Future Fellowship FT170100376.
NMN and GGK acknowledge the support of the Australian Research Council through Discovery Project grant DP170103470.
KS and RJRV acknowledge funding support from NSF Grant 1816462.

%

\vspace{5mm}
\facilities{Keck(KCWI)}









\pagebreak

\bibliography{mrk1486_Te}{}
\bibliographystyle{aasjournal}



\end{document}